\documentclass[12pt]{article}
\usepackage{amsmath,amssymb}
\usepackage{graphics}
\usepackage[dvipdfm]{graphicx}
\usepackage{color}

\makeatletter

\@addtoreset{equation}{section}
\makeatother

\begin{document}
\begin{titlepage}
\begin{flushright}
{\small OU-HET 726/2011}\\
\end{flushright}
\vspace*{1.2cm}

\begin{center}

{\Large\bf 
TeV-scale seesaw with non-negligible  \\ 
left-right neutrino mixings 
} 
\lineskip .75em
\vskip 1.5cm

\normalsize
$^1${\large Naoyuki Haba}, 
$^1${\large Tomohiro Horita},
$^1${\large Kunio Kaneta},
and 
$^2${\large Yukihiro Mimura}

\vspace{1cm}

$^1${\it Department of Physics, 
 Osaka University, Toyonaka, Osaka 560-0043, 
 Japan} \\

$^2${\it Department of Physics, 
 National Taiwan University, 
 Taipei, 10617, Taiwan (R.O.C.)} \\

\vspace*{10mm}

{\bf Abstract}\\[5mm]
{\parbox{13cm}{\hspace{5mm}
%

We suggest natural TeV-scale seesaw 
 with non-negligible left-right neutrino
 mixings as preserving tiny 
 neutrino masses. 
Our analysis is exhibited, without loss of generality, 
 by taking a basis of the neutrino matrices,  
 in which the condition to obtain the left-right 
 mixings is clear. 
We also suggest a flavor symmetry 
 as an underlying theory, which naturally realizes 
 our setup to preserve tiny neutrino masses. 
Our setup can predict a magnitude of
 $\sin\theta_{13}$ in a region 
 of 0.10 to 0.20, depending on the deviation from maximal
 atmospheric neutrino mixing and CP phase. 
We also investigate 
 experimental constraints 
 and phenomenology in our setup.

}}

\end{center}

\end{titlepage}


\baselineskip=16pt

\section{Introduction}

The recent neutrino oscillation experiments 
 gradually reveal the 
 structure of 
 lepton sector\cite{Strumia:2006db, analyses}.  
{}From the theoretical point of view, however, 
 smallness of neutrino mass is still a mystery and 
 it is one of the
 most important clues to find new physics beyond the
 standard model (SM). 
A lot of ideas have been suggested to explain 
 the smallness of neutrino masses comparing to those of quarks and
 charged leptons. 
And type-I seesaw is the simplest mechanism 
  which naturally induces a tiny neutrino masses\cite{Type1seesaw}. 
This mechanism has right-handed neutrinos, and lepton number is 
 softly broken by their Majorana masses. 
The neutrino 
 mass matrix is given by
\begin{equation}
M_\nu = \left(
	\begin{array}{cc}
	0 & M_D \\
	M_D^T & M_R
	\end{array}
\right).
\end{equation}
The seesaw neutrino formula is obtained 
 when $M_D \ll M_R$, where 
 a mass of light neutrino is given by 
\begin{equation}
{\cal M}_\nu^l = - M_D M_R^{-1} M_D^T.
\end{equation}
Usually, with ${\cal O}(10^2)$ GeV Dirac neutrino masses,
 Majorana masses should be super-heavy 
 of ${\cal O}(10^{14})$ GeV. 
Therefore, Majorana nature, such as 
 lepton number violation, is difficult 
 to be observed at any experiments.  
In this situation, 
 mixings between left- and right-handed neutrinos 
 (left-right mixing) are usually negligible, 
 since it is roughly $\sin\theta \sim m_D/m_R = \sqrt{m_\nu/m_R}$,
 where $m_D$ and $m_R$ are the typical Dirac and right-handed Majorana masses,
 respectively for the 1-flavor case, 
 and $m_\nu$ is a light neutrino mass $< {\cal O}(0.1)$ eV.
For example, for $m_\nu = 0.1$ eV and $m_R = 100$ GeV,
 the left-right neutrino mixing is $10^{-6}$. 
Thus, if there is only one generation,
 a tiny neutrino mass direct means 
 a negligible left-right mixing of neutrinos. 
However, the relation between the left-right neutrino mixing
 and the neutrino masses can be drastically changed 
 if we include generation structure
 as it has been pointed out in the literatures\cite{Buchmuller:1990du,He:2009ua,Ma:2009du}. 
An existence of the generation structure can 
 make the left-right mixings non-negligible 
 as preserving a tiny neutrino mass.

In this paper, 
 we suggest natural TeV-scale seesaw 
 with non-negligible left-right 
 mixings as preserving tiny 
 neutrino masses. 
We will show explicitly that 
 the left-right neutrino mixings can be much larger
 than the 1-flavor estimation
 if $(M_D)_{33}$ dominates the other components in $M_D$
 and $(M_R^{-1})_{33} \to 0$
 in 3-flavor case. 
Then, we can obtain 
 sizable left-right mixing as well as 
 suitable tiny neutrino masses. 
Our analysis does not depend on 
 a texture assumption, and 
 one can always take our basis of the matrices
 by rotations of left- and right-handed fields 
 without loss of generality. 
We also suggest a flavor symmetry 
 as an underlying theory,
 which naturally 
 realizes our setup. 
In the setup, a magnitude of
 $\sin\theta_{13}$ can be predicted in a region 
 of 0.10 to 0.20. 
We also investigate 
 experimental constraints 
 and phenomenology in our setup.

\section{2-flavor toy model}

As it is noted in Introduction, the left-right neutrino mixing is
directly related to the light and heavy neutrino masses
in 1-flavor case.
In the multi-generation case, however, the direct relation
is not valid due to new freedom of generation mixings.
In order to illustrate it, 
let us describe the two-flavor toy model. 

Without loss of generality,
one can always take the (1,1) components of Dirac and Majorana mass matrices,
 $M_D$ and $M_R$, to be zero by
 rotations of left- and right-handed 
 fields as 
\begin{equation}
M_D = \left(
	\begin{array}{cc}
	0 & a  \\
	b  & 1
	\end{array}
\right) m,
\qquad
M_R = \left(
	\begin{array}{cc}
	0 & x  \\
	x  & 1
	\end{array}
\right) M.
\end{equation}
Then, 
 the $4\times4$ neutrino mass matrix is 
 given by 
\begin{equation}
M_\nu
= 
\left( 
 \begin{array}{cccc}
   0 & 0 & 0 & a m\\   
   0 & 0 & bm & m \\
   0 & bm & 0 & x M \\
   a m & m & x M & M
 \end{array}
\right).
\label{44}
\end{equation}
Our task is diagonalizing this $4\times 4$ mass matrix.
Here,
 let us show a case with a large left-right mixing 
 as preserving 
 tiny neutrino masses. 
The large left-right mixing can be obtained if $xM \sim m$,
and preserving tiny neutrino masses is possible if $b \ll 1$.
For simplicity to perform the diagonalization, we also assume $a\ll 1$.
(If $a$ is large, rotate the left-handed fields to make the (1,2) element
of the Dirac mass to be zero).


We define a $4\times 4$ unitary matrix $U_{ij}(\theta)$ as
 $(i,i)$ and $(j,j)$ components are $\cos\theta$, 
 $(i,j)$ component is $-\sin\theta$, and
 $(j,i)$ component is $\sin\theta$,
 and the other diagonal (off-diagonal) 
 components are 1 (0), for example, 
\begin{equation}
U_{23}(\theta) = \left( 
 \begin{array}{cccc}
   1 & 0 & 0 & 0 \\   
   0 & \cos\theta & -\sin\theta &0\\
   0 & \sin\theta & \cos\theta  &0\\
   0 & 0 & 0 & 1 
 \end{array}
\right).
\end{equation}
%
%
The $(4,2)$ and $(2,4)$ components in Eq.(\ref{44}) 
 can be rotated away as  
\begin{equation}
M_\nu^\prime = U_{23}(\theta) M_\nu U_{23}(\theta)^T
= 
\left( 
 \begin{array}{cccc}
   0 & 0 & 0 & a m\\   
   0 & -bm \sin2\theta & bm \cos2\theta & 0 \\
   0 & bm \cos2\theta & bm \sin2\theta & x^\prime M \\
   a m & 0 & x^\prime M & M
 \end{array}
\right),
\label{26}
\end{equation}
where $x^\prime = \sqrt{x^2+(m/M)^2}$, 
and $\tan\theta = m/(xM)$.
%
To diagonalize the $4\times 4$ mass matrix,
 (1,4) and (2,3) components in Eq.(\ref{26}) have to be rotated away.
Note that the left-right mixing angles, which rotate 
 away the remaining 
 (1,4) and (2,3) components, 
 are negligibly small due to the tiny light neutrino masses. 
%
%
By those rotations, the (approximate) light neutrino mass matrix is obtained as
in the type-I seesaw formula:
\begin{equation}
{\cal M}_\nu^l = - M_D M_R^{-1} M_D^T
= 
\left(
	\begin{array}{cc}
		0 & - \frac{a b m^2}{xM} \\
		- \frac{abm^2}{xM}& \frac{b^2 m^2}{x^2 M}- \frac{2 b m^2}{xM} 
	\end{array}
\right).
\end{equation}
The second term in the (2,2) component in ${\cal M}_\nu^l$ corresponds to
the (2,2) component in ${\cal M}_\nu^\prime$ : 
 $-2bm^2/(xM) \simeq - b m \sin2\theta$.

Neglecting the tiny angles, which rotate away
 the (1,4) and (2,3) components in Eq.(\ref{26}), 
 we obtain
 the $4\times4$ PMNS (Pontecorvo-Maki-Nakagawa-Sakata) neutrino mixing matrix as
\begin{eqnarray}
U_{\rm PMNS} & \simeq &\left(
	\begin{array}{cc}	
	(U_e^L)^* & 0 \\
	0 & 1			
	\end{array}
\right)
U_{23}(\theta)^T
\left(
	\begin{array}{cc}	
	(U_\nu^L)^T & 0 \\
	0 & (U_\nu^R)^T			
	\end{array}
\right) \\
&=&
\left(
	\begin{array}{cc}	
	(U_e^L)^* \left(\begin{array}{cc} 
			1 & 0 \\
			0 & \cos\theta
		    \end{array}
		\right) (U_\nu^L)^T
& 
	(U_e^L)^* \left(\begin{array}{cc} 
			0 & 0 \\
			\sin\theta & 0
		    \end{array}
		\right) (U_\nu^R)^T 
\\
	\left(\begin{array}{cc} 
			0 & -\sin\theta \\
			0 & 0
		    \end{array}
		\right) (U_\nu^L)^T
& 
	\left(\begin{array}{cc} 
			\cos\theta & 0 \\
			0 & 1
		    \end{array}
		\right) (U_\nu^R)^T 
	\end{array}
\right),
\end{eqnarray}
where $U_e^L$ is a diagonalization unitary matrix
 of the charged lepton mass matrix,
 $U_\nu^L$ is a 
 diagonalization unitary matrix
 of the light neutrino mass matrix ${\cal M}_\nu^l$,
 and $U_\nu^R$ is a diagonalizing unitary matrix of the right-handed neutrino 
 mass matrix.
This $\theta$ angle can be the origin of a sizable left-right neutrino mixing,
 which can be much larger than the 1-flavor case
 i.e., $m_D/m_R \simeq \sqrt{m_\nu/m_R}$.
It can be as large as the experimental bounds.
As shown below in a realistic model, 
 experimental constraints need
 small $\theta$ enough to 
 approximate $\cos\theta\simeq 1$. 
In the usual seesaw with three super-heavy 
 right-handed neutrinos, 
 we only take into account of 
 upper left submatrix with $\cos\theta\simeq 1$, 
 and can neglect left-right mixings,
 where $2\times 2$ active neutrino flavor mixing matrix (we call MNS matrix) 
 is given by 
 $(U^L_e)^*(U_\nu^L)^T$. 
However, in our setup, 
 we can have a non-zero left-right mixing, 
 which contributes phenomenology.

Of course, the above situation is 
 shown from another procedure, i.e.,
 we diagonalize the Majorana mass matrix $M_R$ at first. 
In this basis, 
 neutrino mass matrix is given by 
\begin{equation}
M_\nu^{\prime} = U_{34}(\theta_1) M_\nu U_{34}(\theta_1)^T
= 
\left( 
 \begin{array}{cccc}
   0 & 0 & -s_1 a m& c_1 a m\\   
   0 & 0 & (b c_1 - s_1)m & (bs_1 +c_1)m \\
   -s_1 a m & (bc_1 -s_1)m & M_1 & 0 \\
   c_1 a m & (bs_1+c_1)m & 0 & M_2
 \end{array}
\right),
\end{equation}
where $\tan2\theta_1 = 2 x$,
 $s_1 = \sin\theta_1$, $c_1 =\cos\theta_1$, 
 $M_1 = - M \sec2\theta_1 s_1^2$,
 and $M_2 = M \sec2\theta_1 c_1^2$.
Notice that it satisfies
\begin{equation}
\frac{s_1^2}{M_1} + \frac{c_1^2}{M_2} =0.
\end{equation}
In fact, in the limit of 
 $b\to 0$, 
 $(\frac{s_1^2}{M_1} + \frac{c_1^2}{M_2})(1+a^2)m^2$
 is the seesaw neutrino mass.
The left-right mixings are 
$\theta_{23} \sim s_1 m/M_1$ and $\theta_{24} \sim c_1 m/M_2$,
 and then 
 $\theta_{23}^2 M_1 + \theta_{24}^2 M_2 =0$ is satisfied.
Therefore, one can interpret that 
 this cancellation can realize 
 small seesaw neutrino mass even when the left-right mixing is large.
The cancellation is originated from 
$(M_R^{-1})_{22} = 0$ in the original basis
(in 2-flavor case, it simply means $(M_R)_{11} = 0$).

\section{3-flavor realistic case}

Now let us consider a realistic 3-flavor case. 
The above description is parallel even in the 3-flavor case,
and the key condition to obtain a large left-right mixing
as preserving tiny neutrino masses
 is that 
 $(M_R^{-1})_{33}\to 0$, $(M_D)_{33} \neq 0$
 and $(M_D)_{ij} (j = 1,2)$ are small. 
Therefore, we describe it in the basis where
\begin{equation}
M_R = \left(\begin{array}{ccc}
		0 & 0 & x \\
		0 & y & z \\
		x & z & 1
	    \end{array}
      \right) M.
 \label{MRbasis}
\end{equation}
The Dirac mass matrix is also taken to be
\begin{equation}
M_D = \left(\begin{array}{ccc}
		0 & 0 & a \\
		0 & b & c \\
		d & e & 1
	    \end{array}
      \right) m.
 \label{MDbasis}
\end{equation} 
It should be stressed that these are not a texture assumption.
Without loss of generality, one can always take these basis of the matrices
by rotations of left- and right-handed fields.

The conditions for a large left-right mixing as
 preserving tiny neutrino masses
 are $b,d,e \ll 1$ and $m \sim x M$.
The condition $b,d,e \ll 1$ can be easily constructed
by a well-known flavor symmetry to generate fermion mass hierarchy,
and we will construct a model in the next section.
%

In order to obtain the $6\times 6$ PMNS neutrino mixing matrix,
we need to diagonalize the $6\times 6$ neutrino matrix:
\begin{equation}
M_\nu = \left(
	\begin{array}{cc}
	0 & M_D \\
	M_D^T & M_R
	\end{array}
\right).
\end{equation}
Similarly to the 2-flavor case,
one can rotate away the (3,6) and (6,3) elements
by the mixing angle $\tan\theta = m/(xM)$.
We assume $a,c \ll 1$, for simplicity.
If $a$ and/or $c$ are large, rotate the left-handed fields
to make the (1,3) and (2,3) elements of the Dirac mass matrix to be zero.
Then, one obtains $\tan\theta = \sqrt{1+a^2+c^2} m/(xM)$.
The PMNS neutrino matrix is given in the form:
\begin{equation}
\left( \begin{array}{cc}
 U_{\nu\nu} & U_{\nu N} \\
 U_{N \nu} & U_{NN} 
	\end{array}
\right),
\end{equation}
where $U_{\nu\nu}$ will be the usual MNS neutrino mixing matrix
when left-right neutrino mixing is small.
In the parallel discussion to the 2-flavor case
(neglecting the tiny rotation angles in the limit of $a,b,c,d,e \to 0$),
we obtain
\begin{equation}
U_{\nu\nu} \simeq (U_e^L)^* \left(
		\begin{array}{ccc}
		1 & & \\
		& 1 & \\
		& & \cos\theta
		\end{array}\right)
 (U_\nu^L)^T ,
\end{equation}
and
\begin{equation}
U_{\nu N} \simeq (U_e^L)^* \left(
		\begin{array}{ccc}
		0 & 0 & 0 \\
		0 & 0 & 0 \\
		\sin\theta & 0 & 0
		\end{array}\right) (U_\nu^R)^T
= (U_e^L)^* \left(
		\begin{array}{ccc}
		0 & 0 & 0 \\
		0 & 0 & 0 \\
		p_1 \sin\theta & p_2 \sin\theta & p_3 \sin\theta
		\end{array}\right),
\label{UnuN}
\end{equation}
where $p_i = (U_\nu^R)_{i1}$
and $U_\nu^R$ is a diagonalizing matrix of the right-handed neutrino Majorana mass matrix 
$M_R$, i.e., $U_\nu^R M_R (U_\nu^R)^T = {\rm diag}.(M_1,M_2,M_3)$, 
where the eigenvalues of Majorana masses $M_i$ can be made to be real numbers.
(They all can be made to be positive numbers, 
 but then, (at least) one of $p_i$ has to be pure imaginary
 if all the elements in $M_R$ are real.)
By definition and $(M_R)_{11} = 0$, 
we obtain the relation:
\begin{equation}
p_1^2 M_1 + p_2^2 M_2 + p_3^2 M_3 =0.
\end{equation}


Suppose that the experimentally observed large neutrino mixings in 
the MNS matrix ($U_{\nu\nu}$)
almost come from the charged lepton side,
namely, $(U_e^L)^* \sim U_{\rm MNS}$,
we obtain
\begin{equation}
U_{\nu N} \sim U_{\rm MNS} \left(
		\begin{array}{ccc}
		0 & 0 & 0 \\
		0 & 0 & 0 \\
		p_1 \sin\theta & p_2 \sin\theta& p_3 \sin\theta
		\end{array}\right) ,
\label{UnN1}
\end{equation}
and then,
\begin{equation}
U_{e N_i} \sim U_{e3}\, p_i \sin\theta,
\quad
U_{\mu N_i} \sim U_{\mu 3}\, p_i \sin\theta,
\quad
U_{\tau N_i} \sim U_{\tau 3}\, p_i \sin\theta,
%
\label{UnN1d}
\end{equation}
where $U_{e3}$ corresponds to the 13 neutrino mixing
which is bounded from the CHOOZ experiment 
but ${\cal O}(0.1)$ from the recent T2K/MINOS data 
analyses \cite{Abe:2011sj,Adamson:2011qu},
while $U_{\mu3} \simeq U_{\tau3} \simeq 1/\sqrt2$
from the atmospheric neutrino oscillations.
%
%
In this case, therefore,
\begin{eqnarray}
U_{e N_1} \ll U_{\mu N_1} \simeq U_{\tau N_1}.
\end{eqnarray}
The size of $U_{e N_1}$ is important for the neutrinoless
double beta decay, and
the size of $U_{\mu N_1}$ is important for the LHC experiment.

On the other hand, 
 if the MNS mixing matrix almost comes from the
 neutrino side,
$(U_\nu^L)^T \sim U_{\rm MNS}$,
we obtain
\begin{equation}
U_{\nu N} \sim 
 \left(
		\begin{array}{ccc}
		0 & 0 & 0 \\
		0 & 0 & 0 \\
		p_1\sin\theta & p_2 \sin\theta& p_3 \sin\theta
		\end{array}\right), 
\label{UnN2}
\end{equation}
which shows a mixing between 
 $\nu_\tau$ and $N$. 
In this case, therefore, only 
 $U_{\tau N_i}$ 
 exists, and $U_{e N_i}$ and $U_{\mu N_i}$ are small. 
This is the case where 
 (3,3) component of $M_D$ is dominant ($a,c\to 0$). 
And if (2,3) component dominates, 
 a left-right mixing only exists in $\nu_\mu$ and $N_i$ mixings.

In the same way as in 2-flavor case,
one can start from the diagonalization of $M_R$,
in order to diagonalize $6\times 6$ mass matrix.
In the basis where the right-handed fields are rotated to make $M_R$ diagonal,
$U_\nu^R M_R (U_\nu^R)^T = {\rm diag}.(M_1,M_2,M_3)$,
the Dirac mass matrix 
is calculated as
\begin{equation}
M_D^\prime = M_D (U_\nu^R)^T
= \left(
    \begin{array}{ccc}
      (U_\nu^R)_{13} a & (U_\nu^R)_{23} a & (U_\nu^R)_{33} a  \\
      (U_\nu^R)_{13} c & (U_\nu^R)_{23} c & (U_\nu^R)_{33} c  \\
      (U_\nu^R)_{13}  & (U_\nu^R)_{23}  & (U_\nu^R)_{33}         
    \end{array}
  \right) m,
\end{equation}
in the limit of $b,d,e \to 0$ (for simplicity to show).
In this basis, the left-right neutrino mixings are 
obtained as $U_{\nu N_i} \simeq (U_\nu^R)_{i3}^* m /M_i$.
Note that one can obtain a relation 
$(U_\nu^R)_{i1} (M_R)_{13} = M_i (U_\nu^R)^*_{i3}$ ($i$ is not summed)
from
$U_\nu^R M_R = {\rm diag}.(M_1,M_2,M_3) (U_\nu^R)^*$,
and thus the left-right mixings correspond to the previous ones. 
It is also important to note that
vanishing $(M_R^{-1})_{33}$ in the original basis 
leads the relation:
\begin{equation}
\frac{(U_\nu^R)_{13}^2}{M_1} + \frac{(U_\nu^R)_{23}^2}{M_2}
+ \frac{(U_\nu^R)_{33}^2}{M_3} = 0.
\end{equation}
This relation corresponds to vanishing the seesaw neutrino masses
(in the limit $b,d,e \to 0$)
in the $M_R$ diagonal basis:
\begin{equation}
\frac{(M_D^\prime)_{i1}^2}{M_1} +
\frac{(M_D^\prime)_{i2}^2}{M_2} +
\frac{(M_D^\prime)_{i3}^2}{M_3} = 0.
\end{equation}
This cancellation condition has been studied in several setups
in the literature\cite{Buchmuller:1990du}.
In our choice of the $M_R$ basis, the cancellation
simply corresponds to the vanishing $(M_R^{-1})_{33}$,
and what we need to preserve tiny neutrino mass with 
a large left-right neutrino mixing 
is $b,d,e\to 0$ in the basis.

Here we comment on 
 works in Ref.\cite{He:2009ua,Ma:2009du}. 
In Ref.\cite{He:2009ua}, 
they consider a determinant zero condition for  
 the left-right mixing matrix, $U_{\nu N}$. 
This is a necessary condition to preserve tiny neutrino masses
and 
 they investigate both rank 1 case and rank 2 case,  
 with explicit mixing parameters. 
In Ref.\cite{Ma:2009du},
 they work on a similar analysis as ours,
 but in a different basis with several texture hypothesis.
We take another standing point, that is, 
 the mass matrices originate from flavor symmetry,
 and take explicit forms of 
 Dirac and Majorana mass matrices.

\section{Flavor symmetry}

Let us consider underlying theory 
 which induces the above conditions to preserve
 the tiny neutrino masses with a large left-right neutrino mixing,
 namely, $b,d,e \ll 1$ in the basis given in Eqs.(\ref{MRbasis}) and (\ref{MDbasis}).
The hierarchical structure of the Dirac mass matrix can be easily
constructed in many of the flavor models
as in the other charged fermion mass matrices.
Under the construction,
only one of the elements of the mass matrix, say (3,3) element, 
can dominate others.
As one can find from the discussion in the previous section,
what we need is $(M_R^{-1})_{33} \to 0$
in the basis where $(M_D)_{33}$ dominates other elements of $M_D$.
As we have noted, the form of the matrices
in Eqs.(\ref{MRbasis}) and (\ref{MDbasis}) 
is not a texture hypothesis but a parameterization.
However, 
since the third generation of the right-handed neutrino field
is fixed in the basis where $(M_D)_{33}$ dominates the others in the flavor model
to generate mass hierarchy,
we cannot achieve the basis in Eq.(\ref{MRbasis}) in general by field redefinition
and we need additional symmetry to realize $(M_R^{-1})_{33} = 0$.
A simple example to realize the situation is 
 to introduce a global $SU(2)_F$ flavor symmetry.

We introduce two scalar doublets of the symmetry,
 denoted as $\phi_i^{(n)}$ $(n=1,2)$, where 
 $i$ is an index of the $SU(2)_F$. 
For the matter fields, 
 the 1st and 2nd generations 
 are doublets while 
 the 3rd generations are 
 singlet of the $SU(2)_F$.   
The right-handed neutrino Majorana mass terms
 are induced from VEVs of $B-L$ charged Higgs fields, 
 $\Delta$, $\Delta^\prime$ and $\Delta^{\prime\prime}$. 
The Yukawa interactions are given by 
\begin{eqnarray}
fN_R^3 N_R^3 \Delta 
+\frac{f_1}{M_*}(\phi_i^{(1)}N_R^i)N_R^3 \Delta^\prime
+\frac{f_2}{M_*^2}(\phi_i^{(2)}N_R^i)(\phi_i^{(2)}N_R^i) \Delta^{\prime\prime},
\end{eqnarray}  
and then, the Majorana mass matrix is given as
\begin{equation}
M_R = \left( 
 \begin{array}{ccc}
    f_2\frac{(\phi_1^{(2)})^2}{M^2_*}\Delta^{\prime\prime} &
      f_2\frac{\phi_1^{(2)}\phi_2^{(2)}}{M^2_*}\Delta^{\prime\prime} & 
         f_1\frac{\phi_1^{(1)}}{M_*}\Delta' \\
    f_2\frac{\phi_1^{(2)}\phi_2^{(2)}}{M^2_*}\Delta^{\prime\prime} & 
      f_2\frac{(\phi_2^{(2)})^2}{M^2_*}\Delta^{\prime\prime} &
         f_1\frac{\phi_2^{(1)}}{M_*}\Delta' \\
    f_1\frac{\phi_1^{(1)}}{M_*}\Delta' &
      f_1\frac{\phi_2^{(1)}}{M_*}\Delta' &
         f  \Delta \\
 \end{array}
\right).
\end{equation}
Note that $(\phi^{(1)}N_R)^2$ and 
 $(\phi^{(1)}N_R)(\phi^{(2)}N_R)$ can be forbidden 
 by some extra $U(1)$ or discrete symmetries. 
Thus, we can obtain the mass matrix 
where the determinant of a submatrix is zero ($M_{11} M_{22} - M_{12}^2 =0$) 
and thus
$(M_R^{-1})_{33} = 0$.
In fact, we can take 
 the VEVs of $\phi^{(1,2)}$ as
\begin{equation}
\phi^{(1)}=(\phi', \phi'')^T, \;\;
\phi^{(2)}=(0, \phi)^T,
\end{equation}
by using the rotation of $SU(2)_F$ symmetry,
and then (1,1) and (1,2) ((2,1)) components are zero.


The Dirac Yukawa interactions can be also given by 
\begin{eqnarray}
y \psi_L^{3} \bar{\psi}_R^{3}H
+\frac{1}{M_*}\left(y_1(\phi_i^{(1)}\psi_L^{i})\psi_R^{3}H'
+y_2\psi_L^{3}(\phi^{(1)}_i\bar{\psi}_R^{i})H'\right)
+\frac{y_3}{M_*^2}
(\phi_i^{(2)}\psi_L^{i})(\phi_i^{(2)}\bar{\psi}_R^{i})H^{\prime\prime},
\end{eqnarray} 
where $H$, $H'$ and $H^{\prime\prime}$ are the 
 $SU(2)_L$ doublets. 
Then, one can construct the wanted situation
where $(M_D)_{33}$ dominates other elements 
and $(M_R^{-1})_{33}=0$.

In order to realize the tiny neutrino masses with a large left-right mixing,
we do not need to distinguish $H$ and $H^\prime$ as a construction.
As an option, however, one can also forbid other possible terms
by adopting $H^\prime$:
 $(\phi_i^{(1)}\psi_L^{(i)})(\phi_1^{(i)}\bar{\psi}_R^{(i)})H$, 
 $(\phi_i^{(1)}\psi_L^{(i)})(\phi_i^{(2)}\bar{\psi}_R^{(i)})H$ 
 and  
 $(\phi_i^{(2)}\psi_L^{(i)})(\phi_i^{(1)}\bar{\psi}_R^{(i)})H$ 
 by some symmetries,
and then the productivity of the left-handed neutrino mixings 
is obtained.
At that time, (1,1), (1,2), and (2,1) elements can be zero 
simultaneously for all the lepton Dirac and Majorana mass matrices
without any additional rotation of the charged lepton fields
in the multi-Higgs models \cite{multiHiggs, {Morozumi:2011zu}, GUT}.





By a simple algebra,
if (1,1), (1,2) and (2,1) elements are zero for both
$M_D$ and $M_R$,
those elements are also zero even in the seesaw neutrino mass matrix
${\cal M}_\nu^l = - M_D M_R^{-1} M_D^T$.
If we assume that
 the charged lepton mass matrix is nearly diagonal in the basis
 with negligibly small mixing angles in the diagonalization
 (in this case, the left-right neutrino mixing matrix is 
  given as Eq.(\ref{UnN2})),
 the neutrino mass matrix is given by 
\begin{equation}
{\cal M}_\nu^l \propto  U_{\rm MNS}^* \left(
		\begin{array}{ccc}
		\lambda & 0 & 0 \\
		0 & \epsilon & 0 \\
		0 & 0 & 1 
		\end{array}\right)  
 U_{\rm MNS}^\dagger .
\end{equation} 
Vanishing of the (1,1) and (1,2) components 
 means 
\begin{eqnarray}
&& \epsilon = -e^{-2i\delta}\tan\theta_{13}
(\tan\theta_{13}
 +e^{i\delta}\cot\theta_{12}\sec\theta_{13}\tan\theta_{23}), 
\label{theta-relation1}
\\
&& \lambda=
-e^{-2i\delta}\tan\theta_{13}
(\tan\theta_{13}
 -e^{i\delta}\tan\theta_{12}\sec\theta_{13}\tan\theta_{23}).
\label{theta-relation2}
\end{eqnarray}
In the case of the normal hierarchy, 
\begin{eqnarray}
{\Delta m^2_{12}\over \Delta m^2_{23}}=
{|\epsilon|^2-|\lambda|^2 \over 1-|\epsilon|^2},
\end{eqnarray}
we obtain
\begin{eqnarray}
\frac{\Delta m^2_{12}}{\Delta m^2_{23}}
=
\frac{
4 \sin^2\theta_{13} \csc 2\theta_{12} \tan\theta_{23}
(\sin\theta_{13} \cos\delta + \cot 2\theta_{12} \tan\theta_{23})}{
1- \sin^2\theta_{13}(2+ \cot^2\theta_{12} \tan^2\theta_{23})
 - 2 \sin^3\theta_{13} \cos\delta \cot\theta_{12}
\tan\theta_{23}
}.
\end{eqnarray}
{}From this equation, $\theta_{13}$ mixing is obtained
as a function of $\theta_{23}$, $\theta_{12}$, 
$\frac{\Delta m^2_{12}}{\Delta m^2_{23}}$
and a CP phase $\delta$ in the oscillations. 
The same relation is suggested in the context of 
two-zero texture hypothesis\cite{Fritzsch:2011qv}.
%
%
%
\begin{figure}[tbp]
 \center
  \includegraphics[width=10cm]{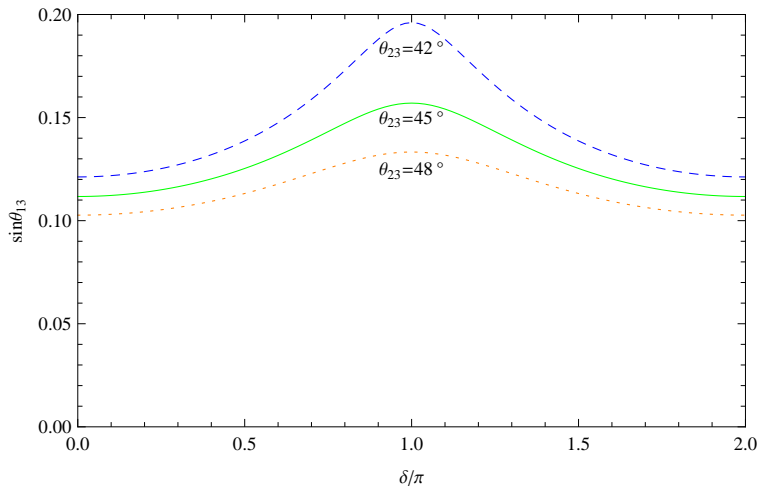}
 \caption{
Magnitude of $\sin\theta_{13}$ depending on $\delta/\pi$ (and also 
 $\theta_{23}$), 
 which is limited in a narrow band. 
}
\label{Fig1}
\end{figure}
We plot $\sin\theta_{13}$ as a function of $\delta$ in Figure 1. 
We can show that 
 the magnitude of $\sin\theta_{13}$ is 
  limited in a narrow band. 
We note that in the case of the inverted hierarchy
 $\sin2\theta_{12}$ needs to be maximal for the solution, which 
 is not suitable for the solar neutrino oscillation 
 experiments.

If the MNS neutrino mixings originate from the charged lepton sector,
on the other hand,
the left-right mixing matrix is given as Eqs.(\ref{UnN1}) and (\ref{UnN1d}).
The charged lepton mass matrix in this case is given by 
 \begin{equation}
M_l =\left(
		\begin{array}{ccc}
		0 & 0 & p \\
		0 & s & q \\
		u & t & r \\
		\end{array}\right),
\end{equation}
where 
 $p\sim q\sim r (\sim m_\tau)\gg t\sim u (\sim m_\mu)\gg s(\sim
 m_e)$.
Note that unitarity matrix which diagonalizes 
 $M_l M_l^\dagger$ is bi-large mixing matrix.
Under the assumption,
%
we obtain 
\begin{equation}
 \tan\theta_{12}\simeq \frac{p}{q}, \quad
\tan\theta_{23}\simeq \frac{\sqrt{p^2+q^2}}{r}, 
\quad
%
\sin\theta_{13} \simeq \frac{m_e}{m_\mu} \sin\theta_{23} \frac{t}{u}.
\end{equation}
This case has no theoretical constrained figure contrary to the 
 above case.

\section{Experimental constraints}

There are experimental constraints on left-right mixings
 of neutrinos, such as neutrinoless double beta decay ($0\nu\beta\beta$) experiment,
 lepton flavor violation (LFV) such as $\mu\to e \gamma$, 
 electroweak (EW) precision data, 
 beam dump experiments, 
 analyses of decays of $\tau$ and $K$, $D$ mesons, 
 and so on. 
These experimental constraints are summarized in Ref.\cite{Atre:2009rg}.
Within 
 those experimental constraints, 
 the mixings with the sterile neutrino 
 can be consistent with 
 neutrino oscillation experiments. 
Below, we will consider the following three cases 
 of the lightest right-handed neutrino mass, i.e., 
 $M_1$ = sub GeV, $M_1=100$ GeV,
 and $M_1=1$ TeV, 
 and show the most stringent experimental constraints 
 on the mixing with each lepton flavor. 

For the electron neutrino,  
 the most stringent 
 constraint of left-right mixing 
 comes from 
 $0\nu \beta \beta$ experiment. 
It is highly sensitive because of its high 
effective luminosity thanks to the observations of many nuclei and
no need for an assumption on another mixings. 
We can obtain the bounds of the mixings  
for a wide range of $M_1$ 
from sub GeV to TeV scale. 
For example, the upper bound on mixing $|U_{eN}|^2$ is 
 $10^{-7} -10^{-8}$, $5 \times 10^{-6}$,
 and $5 \times 10^{-5}$ for $M_1 =$ sub GeV, 
 100 GeV, and 1 TeV,
  respectively. 
Note that the bound 
 for $M_1 \gg 1$ GeV \cite{Atre:2009rg,Benes:2005hn} is given by 
\begin{equation}
\left|\sum_i \frac{U_{e N_i}^2}{M_i}\right| < 5 \times 10^{-5} \text{ TeV}^{-1}.
\end{equation}
In our basis of the neutrino matrices given in the previous sections,
 if the MNS neutrino mixings (almost) come from the neutrino side
 and $(M_D)_{33}$ dominates in the Dirac matrix,
 this condition can be satisfied 
 due to the smallness of $U_{eN_1}$. 
On the other hand, if the MNS mixings (almost) come from
 the charged lepton side, 
 the size of $U_{eN_1}$ is related to $U_{e3}$
 and the parameters must be taken to satisfy this constraint.
In a numerical analysis in the next section,
 we focus on the former case.

For the mixing between $\nu_{\mu}$ and $N$,
the EW precision measurements give the most 
 stringent constraint 
 if $M_1$ is heavier than 100 GeV: 
 $|U_{\mu N}|^2 < 3 \times 10^{-3}$. 
On the other hand, 
 if $M_1$ is sub GeV, the beam dump experiments 
 such as NuTeV and PS 191 provide the constraint: $|U_{\mu N}|^2 < 10^{-6}-10^{-8}$.
The constraint from LFV ($\mu \to e \gamma$) must be taken into account,
 though it is related to the size of $U_{eN}$.
The recent upper limit,
 Br($\mu \to e \gamma) < 2.4 \times 10^{-12}$~\cite{Adam:2011ch}, 
 requires $|U_{eN} U^*_{\mu N}| <1.6 \times 10^{-4}$ $(5.6 \times
 10^{-5})$
 for 100 GeV (1 TeV). 
This is consistent with above mentioned bounds from $0 \nu \beta \beta$ and 
 the EW precision measurements. 
Note that the lighter the intermediate particle becomes,
 the more the amplitude of $\mu \to e \gamma$ is suppressed. 
Therefore, if $M_1$ is sub GeV, 
 the LFV constraint is weaker than other constraints.

Finally, for $|U_{\tau N}|$,  
 the most stringent constraint
 comes from 
 EW precision measurements, $|U_{\tau N}|^2 < 6 \times 10^{-3}$,    
 if $M_1$ is heavier than 100 GeV.   
On the other hand, 
 the analyses of decays of $\tau$ and  
 $K$ and $D$ mesons give the stringent bound as 
 $|U_{\tau N}|^2 < 10^{-3} - 10^{-4}$ \cite{Helo:2011yg}, 
 if $M_1$ is sub GeV. 
We summarize the experimental constrains on the 
 left-right mixing 
 in Table 1.
\begin{table}[tbp]
\small
\begin{center}
\begin{tabular}{l||c|c|c}
mixing & sub GeV & 100 GeV & 1 TeV \\ \hline
$|U_{eN}|^2 $ 
& $10^{-8}$ -- $10^{-9}$ $(0 \nu \beta \beta)$ 
& $5 \times 10^{-6}$ $(0 \nu \beta \beta)$ 
& $5 \times 10^{-5}$ $(0 \nu \beta \beta)$ \\ \hline
$|U_{\mu N}|^2 $ 
& $10^{-6}$ -- $10^{-8}$ (beam dump exp) 
& $3 \times 10^{-3}$ (EW precision) 
& $3 \times 10^{-3}$ (EW precision) \\ \hline
$|U_{\tau N}|^2 $ 
& $10^{-3}$ -- $10^{-4}$ ($\tau$ and meson decay) 
& $6 \times 10^{-3}$ (EW precision) 
& $6 \times 10^{-3}$ (EW precision)
\end{tabular}
\end{center}
\caption{The most stringent experimental constraints (upper bounds) on 
 the left-right mixing according to the lepton flavors.}
\end{table}





\section{Numerical analyses}

Now let us show a numerical analysis in our setup. 
Starting with the following Dirac and Majorana mass matrices, 
\begin{equation}
M_D = \left(
\begin{array}{ccc}
0 & 0 & a \\
0 & b & c \\
d & e & f
\end{array}
\right),
\qquad
M_R = \left(
\begin{array}{ccc}
0 & 0 & x \\
0 & y & z \\
x & z & w
\end{array}
\right).
\label{MR2}
\end{equation}
we obtain the seesaw matrix is
\begin{eqnarray}
{\cal M}_\nu^l = - M_D M_R^{-1} M_D^T
= 
-\left(
\begin{array}{ccc}
0 & 0 & \frac{ad}{x} \\
0 & \frac{b^2}{y} & \frac{be}{y} + \frac{d}{x} \frac{cy - bz}{y} \\
\frac{ad}{x} & \frac{be}{y} + \frac{d}{x} \frac{cy - bz}{y}
&
\frac{e^2}{y} +2\frac{d}{x} \frac{fy - ez}{y} - \frac{d^2}{x^2} \frac{yw-z^2}{y}
\end{array}
\right).
\end{eqnarray}
Note that this seesaw neutrino matrix is rank 1 if $d=0$.

As we have obtained, the left-right neutrino mixing is characterized by
\begin{equation}
(U_\nu^R)_{i1} \sin \theta \simeq (U_\nu^R)_{i1} \frac{f}{x},
\end{equation}
in the limit of $a,b,c,d,e \ll f$.
Similarly to Eqs.(\ref{theta-relation1}) and (\ref{theta-relation2}),
vanishing (1,1) and (1,2) ((2,1)) components gives the relation between mass
and mixing parameters in $U_\nu^R$.
%
We obtain
\begin{eqnarray}
\frac{M_1}{M_2} &=& - e^{2i\delta} \tan^2\theta^R_{13}
- e^{i\delta} \sec\theta^R_{13} \tan\theta^R_{13} \tan\theta^R_{12} \cot\theta^R_{23}, \label{M1M2} \\ 
\frac{M_1}{M_3} &=& - e^{2i\delta} \tan^2\theta^R_{13}
+ e^{i\delta} \sec\theta^R_{13} \tan\theta^R_{13} \tan\theta^R_{12} \tan\theta^R_{23}, \\ 
(M_R)_{13} &=& e^{-i\delta} M_1 \cos\theta^R_{12} \cot\theta^R_{13},
\end{eqnarray}
where
$U_\nu^R M_R (U_\nu^R)^T = {\rm diag}.(M_1,M_2,M_3)$
%
and 
\begin{equation}
U_\nu^R=\left(
\begin{array}{ccc}
1 & 0 & 0 \\
0 & \cos\theta^R_{23} & \sin\theta^R_{23} \\
0 & -\sin\theta^R_{23} & \cos\theta^R_{23}
\end{array}
\right)
\left(
\begin{array}{ccc}
\cos\theta^R_{13} & 0 & e^{-i\delta} \sin\theta^R_{13} \\
0 & 1 & 0 \\
-e^{i\delta}\sin\theta^R_{13} & 0 & \cos\theta^R_{13}
\end{array}
\right)
\left(
\begin{array}{ccc}
\cos\theta^R_{12} & \sin\theta^R_{12} & 0 \\
-\sin\theta^R_{12} & \cos\theta^R_{12} & 0 \\
0 & 0 & 1
\end{array}
\right).
\end{equation}
%
If $x,y,z$, and $w$
 are all real (which means that a phase in $U_\nu^R$ is $0$ or $\pi$),
the equations can be solved analytically and we obtain:
\begin{eqnarray}
\sin^2 \theta^R_{13} &=& 
\frac{M_1^2 (x^2+M_2 M_3)}{x^2 (M_1-M_2)(M_1-M_3)}, 
\label{13} \\
\sin^2 \theta^R_{12} &=& 
\frac{(x^2 + M_1 M_2)(x^2 + M_1 M_3)}{x^2 (M_1 M_2 - M_2 M_3 + M_3 M_1)+M_1^2 M_2 M_3}, 
\label{12} \\
\sin^2 \theta^R_{23} &=& 
\frac{(M_1-M_3)M_2^2 (x^2 + M_1 M_3)}{(M_2-M_3)(x^2 (M_1 M_2 - M_2 M_3 + M_3 M_1)+M_1^2 M_2 M_3)}, 
\label{23} 
\end{eqnarray}
where $x = (M_R)_{13}$.
%
Then, the analytical form of $U_\nu^R$ can be expressed simply as
\begin{eqnarray}
(U_\nu^R)_{11} &=& \sqrt{\frac{M_2 M_3 + x^2}{(M_2-M_1)(M_3-M_1)}}, \\
(U_\nu^R)_{21} &=& \sqrt{\frac{M_1 M_3 + x^2}{(M_1-M_2)(M_3-M_2)}}, \\
(U_\nu^R)_{31} &=& \sqrt{\frac{M_1 M_2 + x^2}{(M_3-M_1)(M_3-M_2)}},
\end{eqnarray}
and $(U_\nu^R)_{11} \simeq 1$ in the limit $M_1 \ll M_2, M_3$.
Since one can obtain $(U_\nu^R)_{11} \simeq 1$ 
even in the case of general complex parameters,
the left-right mixings can be characterized just by $f/x$
in the hierarchical Majorana masses.
Because $f \sim {\cal O} (100)$ GeV, we need $x \sim {\cal O}(10)$ TeV
in order to obtain a left-right mixing ${\cal O}(0.01)$.
Notice that, from the conditions
 $1 \geq \sin^2\theta^R_{23},\ \sin^2\theta^R_{12} \geq 0$,
we obtain $|M_1 M_2| < x^2$ in the limit of $M_3 \to \infty$. 
As a result, $M_2$ is bounded to obtain 
a sizable left-right neutrino mixing for a given value of $M_1$.


The (3,3) element of ${\cal M}_\nu$
can be expressed as
\begin{equation}
-({\cal M}_\nu)_{33} = \sum_{i,j=1,2} (M_D)_{3i} (M_R^{-1})_{ij} (M_D^T)_{j3} +
2 d \frac{f}{x}.
\end{equation}
Remember that the second term comes from the left-right mixing,
and $f/x$ gives the size of it.
Therefore, if the non-negligible left-right mixing ${\cal O}(10^{-2})$
is considered, the natural size of $d$ to obtain a proper neutrino mass 
is ${\cal O}(1)$ eV.
We note that in the TeV-scale seesaw, 
the first term can be comparable to the second term for
much larger size of $d$,
and cancellation can happen between two terms. 
However, such cancellation is unnatural in our context,
and we disregard the possibility.

In the setup of the $SU(2)_F$ flavor model,
one can obtain $d: e = x : z$.
Assuming the relation $e = d z/x$, we obtain
\begin{eqnarray}
{\cal M}_\nu^l = - M_D M_R^{-1} M_D^T
= 
-\left(
\begin{array}{ccc}
0 & 0 & \frac{ad}{x} \\
0 & \frac{b^2}{y} & \frac{cd}{x}  \\
\frac{ad}{x} & \frac{cd}{x} & 2\frac{df}{x}- \frac{d^2 w}{x^2}
\end{array}
\right).
\end{eqnarray}
If a large 23 mixing comes from the charged lepton mass matrix and
a large 23 mixing from the light neutrino matrix is unwanted,
one can realize it by choosing $c\to 0$.
In that case,
the second neutrino mass simply originates from the usual manner of seesaw $b^2/y$. 
The first neutrino mass can be small by choosing $a\to 0$.


We exhibit the numerical examples of $M_D$ and $M_R$ with 
 the three cases, (1): $M_1$ = 100 GeV, (2): 1 TeV, and (3): sub GeV, 
 in Eqs.(\ref{6.16})-(\ref{6.18}). 
%
The (3,3) element of Dirac mass matrix is taken as 100 GeV in all cases.
%
%
The left-right mixing is chosen as 
$U_{\nu N_1} (\equiv \sqrt{|U_{eN_1}|^2+|U_{\mu N_1}|^2+|U_{\tau N_1}|^2})$ = 0.05 
for $M_1$ = 100 GeV and 1 TeV
and $U_{\nu N_1}$ = 0.005 for $M_1$ =  sub GeV
 to satisfy 
 the experimental bounds. 
Because of the relation $U_{\nu N_1} \simeq (U_\nu^R)_{11} f/x$,
fixing $U_{\nu N_1}$ and three Majorana neutrino masses ($M_1$, $M_2$, and $M_3$),
 one can obtain the Majorana mass matrix $M_R$.
If $a$ and $c$ are small as explained above,
the inputs of the light neutrino masses $m_{\nu_3}$ and $m_{\nu_2}$
can determine $b$ and $d$.
In the examples, we choose $a$ and $c$ appropriately not to affect the masses
and the MNS mixings, which come from the charged lepton side.
The elements are all given in the unit of GeV.

\medskip

(1): $M_1 = 100$ GeV,
\begin{eqnarray}
&&M_D = \left(
\begin{array}{ccc}
0 & 0 & 0.53 \\
0 & 5.7 \times10^{-5} & -1.0 \\
4.1 \times 10^{-10} & -7.7 \times 10^{-10} & 100
\end{array}
\right), \nonumber \\
&&M_R = \left(
\begin{array}{ccc}
0 & 0 & 1640  \\
0 & 370 & -3100 \\
1640 & -3100 & 8750
\end{array}
\right).
\label{6.16}
\end{eqnarray}

We choose $M_2 = 1$ TeV and $M_3 = 10$ TeV,
and we obtain $(U_\nu^R)_{11} = 0.82$.

\medskip

(2): $M_1 = 1$ TeV, 
\begin{eqnarray}
&&M_D = \left(
\begin{array}{ccc}
0 & 0 & 0.31 \\
0 & 2.4 \times 10^{-4} & -1.0 \\
3.8 \times 10^{-10} & -1.2 \times 10^{-9} & 100
\end{array}
\right), \nonumber \\
&&M_R = \left(
\begin{array}{ccc}
0 & 0 & 1500 \\
0 & 6670 & -4800 \\
1500 & -4800  & 2800
\end{array}
\right).
\end{eqnarray}

We choose $M_2 = 1.5$ TeV and $M_3 = 10$ TeV,
and we obtain $(U_\nu^R)_{11} = 0.75$.

\medskip

(3): $M_1 = 0.1$ GeV, 
\begin{eqnarray}
&&M_D = \left(
\begin{array}{ccc}
0 & 0 & 0.14 \\
0 & 6.3 \times 10^{-6} & -1.0 \\
5.0 \times 10^{-9} & -3.5 \times 10^{-8} & 100
\end{array}
\right), \nonumber \\
&&M_R = \left(
\begin{array}{ccc}
0 & 0 & 2.0 \times 10^4 \\
0 & 5.0 & -1.4 \times 10^5 \\
 2.0 \times 10^4 & -1.4 \times 10^5 & 1.0 \times 10^7
\end{array}
\right).
\label{6.18}
\end{eqnarray}

We choose $M_2 = 2$ TeV and $M_3 = 10^4$ TeV,
and we obtain $(U_\nu^R)_{11} = 0.99$.

%
%
%


\section{Experimental signatures}

As pointed out in the literature\cite{Atre:2009rg,Datta:1993nm},
the sizable left-right neutrino mixing 
can cause same-sign di-lepton events at hadron colliders
via $W^+W^+$ fusion with a $t$-channel heavy neutrino
exchange
and resonant production of heavy neutrinos
$(q \bar q^\prime \to W \to \ell N_i$).
For ${\cal O}(100)$ GeV Majorana neutrinos,
the resonant production processes dominates,
and 
the same-sign di-lepton events are expected to be observed at the LHC
if the left-right neutrino mixing is just below the current 
experimental bound.

Since the experimental constraints from neutrinoless double beta decay 
is strong, 
it may be difficult to observe like-sign di-electron events.
However, 
one can observe like-sign di-electron events avoiding the 
$0\nu\beta\beta$ constraint
if $(M_R^{-1})_{11} = 0$ (which is $yw - z^2 = 0$ in the notation in Eq.(\ref{MR2})).
As described,
we obtain $U_{e N_i} \simeq (U_e^L)_{13}^* (U_\nu^R)_{i1} \sin \theta$
from Eq.(\ref{UnuN}).
Therefore, the $0\nu\beta\beta$ amplitude 
is proportional to 
\begin{equation}
\frac{(U_\nu^R)_{11}^2}{M_1}+\frac{(U_\nu^R)_{21}^2}{M_2}
+\frac{(U_\nu^R)_{31}^2}{M_3},
\end{equation}
which is equal to the (1,1) element of $M_R^{-1}$.
By definition of the unitary matrix, we have
$M_R^{-1} = (U_\nu^R)^T {\rm diag}.(\frac1{M_1},\frac1{M_2},\frac1{M_3}) U_\nu^R$.

The situation where the $0\nu\beta\beta$ amplitude vanishes
can be constructed
in a model of $SU(2)_F$ flavor symmetry with $U(1)$ or discrete symmetry
as in a similar manner in section 4.
In this case, the 2nd and 3rd generation form $SU(2)_F$ doublet
and the 1st generation is singlet.
If we forbid the $N^1_R N^1_R \Delta$ term by $U(1)$ or discrete symmetry,
one can easily construct the Majorana mass matrix in Eq.(\ref{MR2})
with $yw-z^2=0$:
\begin{equation}
M_R = \left( 
 \begin{array}{ccc}
    0 & f_1 \frac{\phi_1^{(2)}}{M_*} \Delta^\prime 
& f_1 \frac{\phi_2^{(2)}}{M_*}\Delta^\prime \\
f_1 \frac{\phi_1^{(2)}}{M_*} \Delta^\prime
&
f_2\frac{(\phi_1^{(1)})^2}{M^2_*}\Delta &
      f_2\frac{\phi_1^{(1)}\phi_2^{(1)}}{M^2_*}\Delta \\
f_1 \frac{\phi_2^{(2)}}{M_*}\Delta^\prime 
&
    f_2\frac{\phi_1^{(1)}\phi_2^{(1)}}{M^2_*}\Delta & 
      f_2\frac{(\phi_2^{(1)})^2}{M^2_*}\Delta
 \end{array}
\right),
\end{equation}
where one can choose 
a basis of VEVs as 
 $\phi^{(1)} = (\phi^\prime,\phi^{\prime\prime})^T$
 and $\phi^{(2)} = (0,\phi)^T$.
Even if the $0\nu\beta\beta$ amplitude
vanishes,
the cross section of the same-sign di-electron process
does not vanish as long as the heavy neutrino is produced
at the colliders.
In this case, the bound of the $U_{eN_i}$ 
will be obtained by the precision data, and
$U_{eN_i} U_{\mu N_i}^*$ is bounded from
$\mu \to e\gamma$.
We note that $\mu\to e\gamma$ amplitude do not vanish
in general
even if $(M_R^{-1})_{11} = 0$.
If we choose $w$ to be large ($M_3$ large)
satisfying $yw -z^2 = 0$,
the magnitudes of two other eigenvalues of right-handed neutrino masses 
$(M_1,M_2)$ degenerate,
and then, the $\mu\to e\gamma$ amplitude can be cancelled.
In that case, the same-sign $e^\pm e^\pm$, $e^\pm \mu^\pm$, $\mu^\pm \mu^\pm$
events can still have a chance to be observed.

We note that if $y,w\ll x,z$, two heavier eigenvalues ($M_2,M_3$) degenerate.
In this case, however, the $\mu\to e\gamma$ amplitude 
does not necessarily vanish unless $yw - z^2 = 0$ is satisfied.
In a special case where $w = z = 0$,
one obtains $M_1 = y$, $M_2 = -M_3 = x$ (for $y<x$),
and $(U_\nu^R)_{11} = 0$, $(U_\nu^R)_{21} = (U_\nu^R)_{31}$.
The active neutrinos do not mix with the lightest sterile neutrino,
and 
the amplitudes of $0\nu\beta\beta$ and $\mu\to e\gamma$ vanish as a result
of $M_2$, $M_3$ degeneracy.

Here let us comment on leptogenesis. 
If $y w - z^2 = 0$ is satisfied,
two lighter right-handed neutrino masses can be nearly degenerate
for $w \to 0$ ($M_3 \simeq y$) or 
for $w \gg x,y$ ($M_3 \simeq w$).
%
%
This case may work resonant leptogenesis\cite{leptogenesis}, 
where CP asymmetry is enhanced resonantly by degenerate right-handed neutrino masses, 
and the suitable baryon asymmetry is realized even in ${\cal O}(100)$ GeV
Majorana masses.

\section{Summary and discussions}

We have suggested natural TeV-scale seesaw 
 with non-negligible left-right 
 mixings as preserving tiny 
 neutrino masses. 
We have shown explicitly that 
 the left-right neutrino mixings can be much larger
 than the 1-flavor estimation
 if $(M_D)_{33}$ dominates the other components in $M_D$
 and $(M_R^{-1})_{33} \to 0$
 in 3-flavor case. 
Then, we can obtain 
 sizable left-right mixing as well as 
 suitable tiny neutrino masses.

We have proposed a basis of the Dirac and Majorana 
neutrino mass matrices,
which can be always taken by rotations of left- and 
right-handed neutrino fields without loss of generality.
We performed the diagonalization 
to show how the non-negligible left-right neutrino mixing 
can be obtained.
Under the basis we take,
one can clearly find the condition to preserve tiny active 
neutrino masses with an experimentally non-negligible size of 
left-right neutrino mixing.

We have also suggested a flavor symmetry 
 as an underlying theory, 
 which naturally realizes our setup to keep the
 active neutrino masses tiny and obtain
 experimentally accessible left-right neutrino mixings.
In one of our setup of the flavor symmetry,
the magnitude of $\sin\theta_{13}$
is predicted in a region of 0.10 to 0.20.
The numerical prediction depends on the 
deviation from 45 degree of the atmospheric neutrino mixing
and CP phase.
Those parameters can be measured by the neutrino oscillation experiments
if $\sin\theta_{13}$ is in the predicted range,
and therefore, our scenario can be tested.

We have also investigated experimental constraints 
 and phenomenology in our setup
 to study the feasibility of the same-sign di-lepton events at the colliders.
We constructed a model with a flavor symmetry 
 in which the neutrinoless double beta decay amplitude vanishes,
 which gives the most stringent bound for the mixing between electron-neutrino
 and right-handed neutrinos.
In the model, thus, the same-sign di-electron, electron-muon events
 can be allowed to be observed at the LHC, in addition to the di-muon events.




\section*{Acknowledgments}

We thank R.N. Mohapatra 
 for useful and helpful discussions. 
This work is partially supported by Scientific Grant by Ministry of 
 Education and Science, Nos. 20540272, 22011005, 20244028, and 21244036.
The work of Y.M. is supported by the Excellent Research Projects of
 National Taiwan University under grant number NTU-98R0526.


\end{document}